\documentclass[manuscript,screen]{acmart}
\usepackage[utf8]{inputenc}
\usepackage{pdflscape}
\usepackage{afterpage}
\usepackage{capt-of}
\usepackage{booktabs}

\usepackage{tcolorbox}

\usepackage{ifthen}
\newboolean{showcomments}
\setboolean{showcomments}{true} 
\ifthenelse{\boolean{showcomments}}
    {
        \newcommand{\johan}[1]{\textcolor{red}{{\it [Johan says: #1]}}}
        \newcommand{\per}[1]{\textcolor{magenta}{{\it [Per says: #1]}}}
    }
    {
        \newcommand{\johan}[1]{}
        \newcommand{\per}[1]{}
    }

\setcopyright{acmcopyright}
\copyrightyear{2022}
\acmYear{2022}
\acmDOI{10.1145/1122445.1122456}

\acmConference[xxx'22]{}

\begin{document}



\title{Sustaining Open Data as a Digital Common -- Design principles for Common Pool Resources applied to Open Data Ecosystems}

\author{Johan Linåker}
\email{johan.linaker@ri.se}
\orcid{0000-0001-9851-1404}
\affiliation{%
  \institution{RISE Research Institutes of Sweden}
  \city{Lund}
  \country{Sweden}
}\author{Per Runeson}
\email{per.runeson@cs.lth.se}
\orcid{0000-0003-2795-4851}
\affiliation{%
  \institution{Lund University}
  \streetaddress{P.O. Box 118}
  \city{Lund}
  \country{Sweden}
}

\begin{abstract}
\emph{Motivation.} Digital commons is an emerging phenomenon and of increasing importance, as we enter a digital society. Open data is one example that makes up a pivotal input and foundation for many of today's digital services and applications. Ensuring sustainable provisioning and maintenance of the data, therefore, becomes even more important.
\emph{Aim.} We aim to investigate how such provisioning and maintenance can be collaboratively performed in the community surrounding a common. Specifically, we look at Open Data Ecosystems (ODEs), a type of community of actors, openly sharing and evolving data on a technological platform.
\emph{Method.} We use Elinor Ostrom's design principles for Common Pool Resources as a lens to systematically analyze the governance of earlier reported cases of ODEs using a theory-oriented software engineering framework.
\emph{Results.} We find that, while natural commons must regulate consumption, digital commons such as open data maintained by an ODE must stimulate both use and data provisioning. Governance needs to enable such stimulus while also ensuring that the collective action can still be coordinated and managed within the frame of available maintenance resources of a community. Subtractability is, in this sense, a concern regarding the resources required to maintain the quality and value of the data, rather than the availability of data. Further, we derive empirically-based recommended practices for ODEs based on the design principles by Ostrom for how to design a governance structure in a way that enables a sustainable and collaborative provisioning and maintenance of the data. 
\emph{Conclusion.} ODEs are expected to play a role in data provisioning which democratize the digital society and enables innovation from smaller commercial actors. Our empirically based guidelines intend to support this development.
\end{abstract}

\ccsdesc[500]{Software and its engineering~Open source model}

\keywords{Open Data, Data Ecosystems, Digital Commons, Ostrom, Sustainability, Common Pool Resources}

\maketitle
\section{Introduction}

With the emergence of a digital society, the concept of commons has been applied to digital phenomena with applications ranging, from Open Source Software (OSS)~\cite{schweik2012internet,  benkler2014between} to open data~\cite{RuhaakMozillaCommons2021, CoyleValue2020} and online projects such as Wikipedia~\cite{forte2009decentralization, viegas2007hidden}. A main reason for this relates to the parallel between the sustainability of digital commons~\cite{morell20148}, and the phenomenon referred to as ``the tragedy of the commons'', where individuals act in self-interest and over-utilize the common resource. The ``tragedy'' is commonly exemplified through an open pasture where the rational herder out of self-interest aims to maximize his (momentary) benefit from the pasture by adding more animals, which in the end leads to over-grazing and thereby reducing the benefit that can be gained from the pasture by any other herder~\cite{Hardin1968}. 

The open pasture can in this sense be considered a Common Pool Resource (CPR), implying a resource system where it is difficult (or costly) to exclude other actors from benefiting from the use of its resource units (i.e., grass in the herding example)~\cite{Ostrom90}. A second trait, also in the open pasture example, is that a CPR is subtractable in the sense that over-utilization by some actors may reduce the benefit to be obtained from others if the system is not maintained.

To manage the tragedy, and the will to ``free-ride'' on the cost of others, researchers have argued for central regulation or privatization of the CPR being the only options forward. However, as explained by Ostrom~\cite{Ostrom90}, CPRs can very well be sustained (at least in small and local communities). From her work, design principles, that characterize robust institutions for managing CPRs, have emerged~\cite{Ostrom90}. These have later been validated and extended by Cox et al.~\cite{Cox2010}, empirically confirming them being well supported. 

Preliminary proposals of how to map Ostrom's principles to digital commons have been proposed~\cite{CoyleValue2020, RuhaakMozillaCommons2021, forte2009decentralization, rozas2017self}. However, when mapping theory from general commons to digital commons, it is worth noting a fundamental difference. Digital commons, such as software and data, can be consumed by multiple consumers at the same time without reducing the benefits extracted by each consumer as the marginal production cost is close to zero (copying data or software~\cite{benkler2014between}), implying that digital commons are non-rival~\cite{CoyleValue2020, o2003guarding}. As a consequence, considering the ecosystem in which the digital common is shared and maintained, the value increases -- rather than the opposite -- since having more members raises the probability that the ecosystem stays alive and the common resource is maintained~\cite{iansiti2004keystone}. 

The limitations are rather in labor and computing resources to maintain the commons~\cite{curto2022sustainability}, and as in Hardin's tragedy and following streams of literature, free-riding is a large issue. Limited or no maintenance of the digital common may result in vulnerabilities being introduced, dependencies breaking, machine learning training-sets becoming invalid, or business value being lost leading to users abandoning the common to further deteriorate as a consequence~\cite{curto2022sustainability}. 


In this paper, we investigate how the value of a digital common~\cite{morell2010governance} in the instance of open data~\cite{RuhaakMozillaCommons2021, CoyleValue2020} (i.e., \textit{not open hardware or software}), can be sustained by its users and producers organized in a surrounding ecosystem~\cite{benkler2014between, morell20148}. Specifically, we explore how Ostrom's design principles may be applied in the context of ODEs to design a governance structure that sustains the development, sharing, and quality of the open data, along with its boundary resources, while still enabling open use and appropriation similar as to Wikipedia~\cite{viegas2007hidden, forte2009decentralization} and OSS commons in general~\cite{morell20148, schweik2012internet, rozas2017self}. For each of the design principles, we analyze previously reported cases of open data~\cite{RunesonJSS21,viegas2007hidden} and open software ecosystems~\cite{forte2009decentralization, rozas2017self}, and observe how their governance has been set up to enable joint maintenance of the data as well as its boundary resources. The analysis is then validated against the earlier reported practice proposals of how to map Ostrom's principles to digital commons~\cite{CoyleValue2020, RuhaakMozillaCommons2021}.

We find, in the case of ODEs, that the governance is designed to \emph{promote participation and encourage contributions}, rather than monitor and sanction usage of a common, as the case with the original design principles. Specifically, we propose for each principle, a recommendation for orchestrators or members of ODEs in how they may design their governance in such a way that it promotes the sustainability of open data and the potential value it can provide in existing and future ecosystems.

\section{Background and related work}

\subsection{Open Data Ecosystems}
\label{subsect:rw:OGD}
The ecosystem lens, stemming from the biological setting, has been used both on business~\cite{iansiti2004keystone} and technical levels~\cite{jansen2009sense} to describe how actors collaborate in a common context towards a common goal while benefiting from the joint efforts more than what they would in isolation. Software ecosystems, including OSS, is a well-explored area where actors collaborate on the development and maintenance of a common software platform, that actors in turn can reuse and extend in a common market or type of use case~\cite{alves2017software}. Recently, the lens had been adopted to data ecosystems as well~\cite{oliveira2019investigations}.

Following work by Runeson et al.~\cite{RunesonJSS21, LinakerJeDEM21}, we define an Open Data Ecosystem as a networked \emph{community of actors} (organizations and individuals), which base their relations to each other on a \emph{common interest}. This interest is underpinned by a \emph{technological platform} that enables actors to process data (e.g., find, archive, publish, consume, or reuse) as well as to foster innovation, create value, or support new businesses. Actors \emph{collaborate on the data and boundary resources} (e.g., software and standards), through the exchange of information, resources, and artifacts.

The technological platform, that enables and facilitates the collaboration, is maintained and facilitated by a platform provider, who usually is the main actor that initiated the ecosystem and orchestrates the overarching collaboration~\cite{LinakerJeDEM21}. This actor is commonly constituted by a public sector organization, as sharing of data from and between private actors is still an emerging phenomenon~\cite{oliveira2019investigations}. In cases where the data being shared openly is either produced or collected by a public sector organization, the data is commonly referred to as \textit{open government data} (OGD)~\cite{LinakerJeDEM21}.

The ecosystem actors may be categorized based on their importance for the ecosystem at large and the influence, formal or informal, that usually follows regarding the platform and collaboration. This can be visualized using Nakakoji et al.'s onion model, where those actors in the layers closest to the center have a higher level of influence compared to those further out who are usually more passive or inactive~\cite{nakakoji2002evolution}. The center is made up of the platform provider or group of actors (commonly referred to as keystones) who are leading the decision-making.


On a functional level, actors in an ecosystem can take on one or multiple roles, e.g., data providers, data brokers, infrastructure and tool providers, service providers, application developers, and application users~\cite{immonen2014requirements, kitsios2017business}. They interact in what can be referred to as value networks, where data is shared, enriched, and used as input to new and improved products, services, or use cases that create value for the actors~\cite{lindman2015business, immonen2014requirements}. However, incentives may not necessarily be tied to the data, but rather the knowledge and boundary resources shared as a consequence of the collaboration~\cite{RunesonJSS21}.

\subsection{The emergence of digital commons}
Commons is a broad term, seldom explicitly defined~\cite{hess2008mapping, jullien2020digital}. According to Hess and Ostrom, commons conveys a \textit{``notion of shared ownership, participation, and responsibility''}~\cite{hess2006framework}. Hess later proposes the definition of a common as a \textit{``resource shared by a group where the resource is vulnerable to enclosure, overuse, and social dilemmas. Unlike a public good, it requires management and protection in order to sustain it.''} Underpinning the definition, is a mapping of what types of commons that has emerged, where Hess distinguishes between the traditional commons (such as fisheries, forests, and grazing lands), and other types as that of knowledge commons and its subset of \textit{online mass collaboration projects} (including Wikipedia, and OSS in general)~\cite{hess2008mapping}. 

The traditional type of commons aligns well with what Ostrom considers as CPRs, i.e., a resource system that is large enough to be non-exclusive, yet subtractable in terms of resource units within the system~\cite{Ostrom90}. Knowledge commons on the other hand, and online mass collaboration projects specifically, expand the understanding of what a common may imply. Benkler characterizes this expansion with the distinction between bounded and open commons~\cite{benkler2014between}. The \emph{bounded} common resource is appropriated exclusively and governed within a community as with traditional commons. The \emph{open} common resource has no boundaries in terms of appropriation and is open access (in full or part) to the public as with knowledge commons. Schweik and English make an aligning distinction between environmental and OSS-based commons, where the focus of the community in the latter is on co-creating a public good for anyone to use rather than overuse as in the former~\cite{schweik2012internet}, aligning with e.g., Jullien and Roudaut~\cite{jullien2020digital}. Vi{\'e}gas et al.~\cite{viegas2007hidden} distinguishes between on-line and off-line communities characterized by the works of Benkler~\cite{benkler2014between} and Ostrom~\cite{Ostrom90} respectively.

Benkler describes the open and OSS-based commons as a product of commons-based peer-production~\cite{benkler2005common}, i.e., as the case when a publicly available common is used as input to a collaborative and coordinated production process between a number of individuals, and the output from the process is released back to the originating common~\cite{benkler2002coase}. Fuster Morell describes the communities, in which this peer-production takes place, as \textit{Online Creation Communities} where individuals collaborate through an online platform in creating and sharing a CPR~\cite{morell20148}. These communities, exemplified through Wikipedia and OSS, are open to anyone wishing to use or contribute to the common development of the CPR.

These characteristics are highlighted by Fuster Morell~\cite{morell2010governance} in her definition of digital commons as \textit{``information and knowledge resources that are collectively created and owned or shared between or among a community and that tend to be non-exclusive, that is, be (generally freely) available to third parties. Thus, they are oriented to favor use and reuse, rather than to exchange as a commodity. Additionally, the community of people building them can intervene in the governing of their interaction processes and of their shared resources.''}

\subsection{Governance of digital commons related to software and data}
Governance, as with commons, also comes with multiple definitions. Markus defines it \textit{in the context of OSS} as \textit{``the means of achieving the direction, control, and coordination of wholly or partially autonomous individuals and organizations on behalf of an OSS development project to which they jointly contribute''}~\cite{markus2007governance}. De Laat in turn considers governance as configurations primarily based on democratic or autocratic principles~\cite{de2007governance}. Configurations that can be nuanced even further~\cite{de2013evolution, germonprez2014collectivism}, evolve with time~\cite{o2007emergence}, and even co-exist~\cite{shaikh2017governing}. In a similar context, Alves et al.~\cite{alves2017software} define governance mechanisms as \textit{``managerial tools of participants in software ecosystems, i.e., orchestrators and platform extenders that have the goal of influencing an ecosystem's health. Ecosystems are healthy when they exhibit longevity and propensity for growth.''} A definition further contextualized in the case of data ecosystems by Oliviera et al.~\cite{oliveira2019investigations}.

The connection between governance and health aligns with the intention of Ostrom regarding her design principles of a governance structure that enables a sustainable CPR, i.e., a scenario where its resources are neither depleted nor degraded~\cite{Ostrom90}. The applicability of the design principles in terms of digital commons such as open data and OSS has, however, only been explored to a limited extent (with notable exceptions~\cite{viegas2007hidden, forte2009decentralization, rozas2017self, RuhaakMozillaCommons2021, CoyleValue2020}). One reason for this may be that the CPR framework aligns more with the traditional commons, rather than knowledge and digital commons~\cite{hess2008mapping}. However, as hinted by Benkler~\cite{benkler2014between}, work on CPR may still provide input to the governance within the community where the commons-based peer-production takes place.

This is illustrated by Vi{\'e}gas et al.~\cite{viegas2007hidden} and Forte et al.~\cite{forte2009decentralization} who apply Ostrom's design principles to the governance of Wikipedia. Forte et al. state that the principles fit as the \textit{``[Wikipedia] community is not only managing a resource, it is striving to encourage collaboration and cooperation among volunteers''}~\cite{forte2009decentralization}. They found that the governance is upheld and facilitated through social norms developed by the community through time~\cite{forte2009decentralization}, aligning with Ostrom's suggestion that norms have a higher impact and long-term acceptance than external rules imposed on a community~\cite{ostrom2000collective}.

Rozas~\cite{rozas2017self} performed a similar study on self-organization carried out in the Drupal OSS community. Rozas used the design principles by Ostrom to describe and capture related organizational changes in the Drupal OSS community as it has grown and adopted a decentralized governance and decision-making process. The connection between OSS commons and CPRs is further explored by O'Mahony who confirms that the two share certain characteristics in terms of regulating the balance between provisioning and appropriation~\cite{o2003guarding}, where licences are central in the digital domain. O'Mahony specifically investigates how copyleft license obligations are enforced to make (especially) commercial appropriators contribute back modifications or additions to the OSS (i.e., the common) to preserve its quality.


In terms of open data, Coyle et al. make the analogy of how an upstream farmer may need to share water with a downstream community to improve harvest so that everyone can benefit from~\cite{CoyleValue2020}. According to Coyle et al., \textit{``the holder of data may need to sacrifice some private economic benefits by sharing data to unlock potentially much larger benefits for their sector or supply chain.''} In their report, Coyle et al. further highlight how Ostrom's design principles for CPRs can guide the regulation and governance of the data, by contextualizing the principles from a data economy perspective. Ruhaak et al. make a similar translation of Ostrom's design principle to the context of data commons~\cite{RuhaakMozillaCommons2021}.

\section{Research Approach}
\subsection{Methodology}



The research is a deepened and refocused analysis of existing literature and empirical studies of ODEs~\cite{RunesonJSS21, forte2009decentralization,rozas2017self}, using Ostrom’s design principles as an analytical lens~\cite{Ostrom90}. For each of the design principles, we contrast previously reported cases of digital commons and how the governance has been set up to enable joint maintenance of the data, as well as its boundary resources. The analysis is then contrasted with the earlier reported proposals of how to map Ostrom’s principles to open data as a digital common. 
We specifically address the questions of (RQ1) \textit{how Ostrom’s design principles may be applied in the context of ODEs to help sustain the shared data},  and (RQ2) \textit{what the fundamental differences are between natural and digital commons}.


\begin{figure}
    \centering
    \includegraphics[width=0.8 \columnwidth]{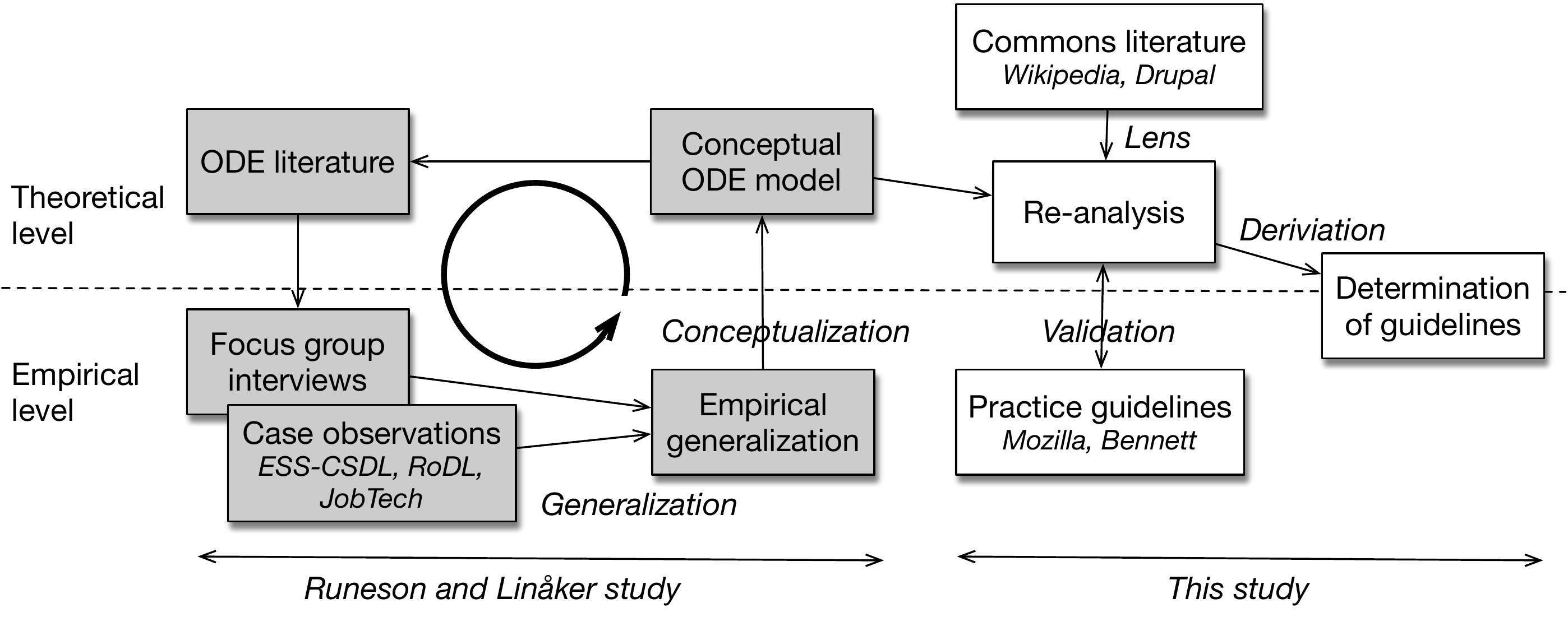}
    \caption{Overview of research method, based on Stol and Fitzgerald~\cite{StolTheory2015}. Grey boxes refer to Runeson and Linåker's study~\cite{RunesonJSS21} and white boxes to the current one.}
    \label{fig:method}
\end{figure}

Conceptually, we follow Stol and Fitzgerald’s theory-oriented software engineering process~\cite{StolTheory2015}, as outlined in Fig.~\ref{fig:method}. In Runeson et al's previous work, focus group interviews were conducted, and observations were made in three ODE practice cases (ESS--CSDL, RoDL, and JobTech), which were conceptualized into a general conceptual model of ODE~\cite{RunesonJSS21}. The model consists of four higher-level themes: \emph{value, intrinsics, governance}, and \emph{evolution}. Of these, \emph{governance} is of particular interest for this study.

Using Ostrom's eight design principles for commons as a lens, we revisit the analysis and the underpinning empirical data of the conceptual ODE model and its governance theme specifically. For each of the eight principles, we firstly contextualize other researchers' interpretations and application of the principles, both in terms of a) case studies of Wikipedia~\cite{forte2009decentralization,viegas2007hidden} and Drupal~\cite{rozas2017self}, and b) two sets of existing practice guidelines, related to Ostrom's guidelines and open data as a digital common; one set by Coyle et al.~\cite{CoyleValue2020} and one published by the Mozilla foundation~\cite{RuhaakMozillaCommons2021}. However, as the derivation of these guidelines is not transparently reported, we consider them to be directly practice-based rather than theoretically founded (labeled as a ``shortcut'' by Stol and Fitzgerald~\cite{StolTheory2015}). Secondly, each principle is contextualized in terms of ODE governance, followed by an analysis and comparison between the ODE contextualization and that of the related work. Thirdly, we derive recommendations for how governance of ODEs can be performed in a way that promotes and enables sustainable maintenance of the digital commons overseen by an ODE. 


\subsection{Cases}
In our analysis, we consider three cases from our previously reported work~\cite{RunesonJSS21} where further details are available.

\textit{Case 1} concerns the The Road Data Lab (RoDL), a joint industry--academia project focused on sharing and collaborating on image and sensor data related to roads. The goal is to provide a basis for machine-learning-based applications related to autonomous driving, but also to explore challenges related to the sharing of data and how these can be addressed.

\textit{Case 2} regards the ESS Control System Data Lab (ESS--CSDL), part of The European Spallation Source (ESS) consortium, which is responsible for managing data from the control systems of the consortium's multi-disciplinary research facility. The ecosystem's scope is to share and collaborate on alarm data, stemming from the control systems and provide a basis for joint innovation and knowledge sharing on how the data can be leveraged to improve the security and operations of the research facility (and related contexts).

\textit{Case 3} focus on JobTech Development (JobTech), an ODE initiated and orchestrated by the Swedish Public Employment Service focused on sharing and collaborating on data that can help improve the digital match-making between job seekers and employers in Sweden. Data includes job ads, personal resumés, and a taxonomy of skills and work titles.

\section{Results and Analysis}
\label{sec:results}

\subsection{Principle 1: Clearly defined boundaries}

\textit{``Individuals or households who have rights to withdraw resource units from the CPR must be clearly defined, as must the boundaries of the CPR itself''}~\cite{Ostrom90}.

\subsubsection{Related interpretations on Digital commons:}
Forte et al. note that the principle, in terms of traditional commons, is focused on defining who has the right to withdraw resource units from a common that is consumable~\cite{forte2009decentralization}. In terms of Wikipedia, which output can be classified as open-access, the principle is re-interpreted to define who is included and has the right to contribute to the commons-based peer-production. This is considered especially important since anyone is eligible to participate without being familiar with the processes of the peer-production. 


This interpretation was further adopted by Rozas in the case of the Drupal OSS community, who noted an increased formalization and definition of these boundaries as the community grew~\cite{rozas2017self}. The interpretations of Rozas and Forte et al., hence, mainly focus on who can \emph{contribute} to the common, and as a consequence with whom to communicate norms and processes related to the peer-production. In essence, this aligns rather well with the intention of Ostrom who considers the principle as a means to help members of a community know \textit{``who is in and who is out of a defined set of relationships and thus with whom to cooperate''}~\cite{ostrom2000collective}.

From a data economy perspective, provided by Coyle et al., the principle also covers the appropriation aspects as highlighted in the original definition: \textit{``Clarity on the rights of different entities to control, access, use and share data''}~\cite{CoyleValue2020}. Ruhaak et al. align in their interpretation concerning data commons as the boundaries should help to \textit{``determine who can contribute, access, and use the data resource or make decisions about it. [The boundaries] also [help] determine the shape and context of the data resource itself''}~\cite{RuhaakMozillaCommons2021}

\subsubsection{Contextualization on ODEs:}
Actors within an ODE base their interrelations on a common interest with the rationale that working together and sharing resources creates more value than standing alone. The common interest, which implicitly leads to defining the boundaries, is the basis for a common vision and purpose for the ODE, and thereby what type of resources should be shared and collaborated on, i.e., the type of data and related boundary resources. The boundaries and vision for the ecosystem is commonly defined by the platform provider and keystone actors closest to the core of the ecosystem (see Section~\ref{subsect:rw:OGD}). 

Among the cases studied, we could differentiate between three types of driving forces among these core groups of actors that help to shape the common vision and way forward.
One type concerns the \emph{business-driven} ODEs where there is an emphasis on creating commercial value and opportunities for companies involved in the ecosystem (exemplified by ESS--CSDL and RoDL). \emph{Public-driven} ODEs (as in the case of JobTech) are a second type where the main focus is on creating value for the society at large. In the third type, the \emph{community-driven} ODEs, the driving force is not necessarily limited to a public or business-centered incentive, but may also include personal or group-wise incentives for individuals and organizations, either co-located or decentralized. Although not studied as a case, OpenStreetMap may be considered as an example. Among the three types mentioned, however, it should be noted that they are not mutually exclusive. An ODE may very well have commercial goals while the overall purpose can be classified as public-driven.

\subsubsection{Analysis and recommendation:} 
As in all related cases, the output of the commons-based peer production in the ODEs is available as open access. Yet, there is still value in defining who will benefit from and use the data as this can be considered part of the ecosystem's common goal and vision. While the cases of Wikipedia~\cite{forte2009decentralization} and Drupal~\cite{rozas2017self} mainly consider boundaries in terms of contributing and engaging in the peer-production, Coyle et al.~\cite{CoyleValue2020} and Ruhaak et al.~\cite{RuhaakMozillaCommons2021} also consider the use and appropriation perspective of the data as important to define as well. In an ODE, the boundaries are implicitly defined by the common vision and goal of the ecosystem. Especially in public- and business-driven ODEs, this is set in the beginning by the platform provider and potentially also keystone actors. However, to build a sustainable and attractive ecosystem, the voices of the actors also need to be considered, as often done implicitly in community-driven ODEs like Wikipedia and Drupal where consensus is a driving norm for decision-making~\cite{viegas2007hidden, forte2009decentralization}.

\textit{Recommendation \#1 for ODEs:} 
Boundaries of the ecosystem and its underpinning common (e.g., in terms of control, access, use, and sharing of data) follow as a consequence of the overarching vision of the ecosystem, which should be inclusive and shaped with the input of the whole ecosystem. 

\subsection{Principle 2: Congruence between appropriation and provision rules and local conditions}

\textit{``Appropriation rules restricting time, place, technology, and/or quantity of resource units are related to local conditions and to provision rules requiring labor, material, and/or money.''}~\cite{Ostrom90}

\subsubsection{Related interpretations on Digital commons:}
In the case of Wikipedia, the growing community has created a decentralized structure of local groups that are responsible for distinct WikiProjects. Grouping within these projects form what can be referred to as local jurisdictions, \textit{``within which local leadership, norms, and standards for writing are agreed upon by editors familiar with a particular topic''}~\cite{forte2009decentralization}. Forte et al., in their reinterpretation, highlight the need to consider these aspects even though central editing and policy guidelines from Wikipedia must still be considered~\cite{forte2009decentralization}. They further highlight the perspective that these local norms may change over time.


The need for a large and decentralized community such as Wikipedia to consider local conditions is confirmed by Vi{\'e}gas et al. who highlight that \textit{``instead of relying on ''one-size-fits-all'' regulation, rules must be intimately associated with the particularities of the resources they regulate''}~\cite{viegas2007hidden}. Rozas further notes, in the case of the Drupal OSS community, an \textit{``emergence of autonomous spaces with regards to local decision-making in contributed projects and local institutions to avoid a “one-size-fits-all” regulation''}~\cite{forte2009decentralization}. 

Coyle et al. interpret this principle slightly differently, focusing on transparency and understanding, rather than adaptations to local conditions: \textit{``Transparency and better understanding of both rights and how value from data returns to people and organisations''}~\cite{CoyleValue2020}. 

\subsubsection{Contextualization on ODEs:}
The diversity of actors and the relationships in-between them affect the conditions and form how data and related resources are shared and shaped through the joint contributions (cf. \textit{``labor, material, and/or money''}~\cite{Ostrom90}). Benefits need to be weighed against the potential risks and costs of sharing data or any boundary resource~\cite{Linaker20}. Companies, for example, commonly associate the risk of giving away differentiating value to competitors through data sharing. A more general, but also overlapping risk concerns that of sharing integrity and confidentiality-related information.

To manage these risks, licenses can be defined to balance how the data is permitted to be shared while still enabling opportunities for individual and collective value creation. Among the cases studied, we have noted four ways in which the data can be transformed before publication to address concerns and perceived risks among ecosystem actors (aligning with e.g.,~\cite{Enders2019}): 1) Adapting the \emph{currentness} of the data, e.g., by only sharing historical data when the real-time data is considered sensitive. 2) Adapting the \emph{level of processing} of the data, e.g., by only providing the raw data when processing and enrichment are considered as differentiating steps. 3) Adapting the \emph{granularity} of the data, e.g., by abstracting, anonymizing, or removing certain fields in the published data. 4) Adapting the \emph{level of openness} of the data, e.g., if it should be shared internally within the ecosystem or openly with an open data license.

\subsubsection{Analysis and recommendation:}
The related studies of Wikipedia~\cite{viegas2007hidden, forte2009decentralization} and Drupal~\cite{rozas2017self} focus on the need to adapt rules, especially for the contribution and collaborative aspects, to local needs. Both cases have very large communities where sub-projects have emerged as a consequence, hence the specific need to consider local conditions. The studied cases of ODEs, however, are rather small and do not have this specific concern in their current stage of evolution. 
Still, they need to define feasible licenses for use and sharing, to ensure that actors in the ecosystem feel that perceived risks are managed and that the potential for value creation persists. As highlighted by Coyle et al.~\cite{CoyleValue2020}, these rules and processes need to be transparent and generally understood, e.g., in how they address the concerned risks.

\textit{Recommendation \#2 for ODEs:} Licenses and processes related to consumption and contribution to the common (i.e., data and related boundary resources) should address perceived risks while persisting potential for value creation for the members of the ecosystem. As the ecosystem evolves, local needs must be considered and addressed accordingly.

\subsection{Principle 3: Collective-choice arrangements}

\textit{``Most individuals affected by the operational rules can participate in modifying the operational rules.''}~\cite{Ostrom90}

\subsubsection{Related interpretations on Digital commons:}
Forte et al. generally align in their interpretation of this principle: \textit{``In order to best accomplish the congruence called for in principle 2, principle 3 suggests that people who are affected by the rules of the community can participate in changing them.''}~\cite{forte2009decentralization}
Vi{\'e}gas et al. further add that \textit{``the cost of altering rules should be kept low''}~\cite{viegas2007hidden}. In terms of Wikipedia, both aforementioned studies describe a decision-making process that is generally based on consensus building among those interested~\cite{forte2009decentralization, viegas2007hidden}. Rozas notes in the case of Drupal that there is a continuous ongoing process in the community in terms of inventing \textit{``ways to increase participation in the elaboration of rules by those affected by them''}~\cite{rozas2017self}. A similar interpretation is echoed by Ruhaak et al.~\cite{RuhaakMozillaCommons2021}.

\subsubsection{Contextualization on ODEs:}
Rules and rights for using and sharing data, as well as contributing to its evolution are ultimately decided by the platform provider. However, the platform provider must still recognize and consider the opinions of the ecosystem actors, especially the keystones as they are pivotal for the success and health of the ecosystem (see principles 4--6 for discussion on the mutual monitoring, sanctioning, and conflict-resolution mechanisms). 

Among our cases, we have noted both formal and informal means for setting up such governance structures and thus we differentiate between three types of ecosystem governance structures.
In \emph{organization-centric} ODEs (e.g. ESS--CSDL and JobTech), the governance is concentrated on one single organization, also constituting the platform provider. Here, the platform provider needs to build relationships and means for the ecosystem to influence, e.g., what data is shared. In \emph{consortium-based} ODEs (e.g RoDL), several actors (often keystones) share the role of the platform provider through a joint organization. Here, governance is set up more formally with e.g., statutes and committees that enable collective decision-making. In \emph{community-based} ODEs (e.g. OpenStreetMap), the role of a platform provider is taken on by a community at large where governance is decentralized directly to the actors within the ecosystem whom themselves decide on how to organize the decision-making.


\subsubsection{Analysis and recommendation:}
All related cases are clear in that those affected by a decision should be able to influence the decision. In Wikipedia~\cite{forte2009decentralization} and Drupal~\cite{rozas2017self}, this is a consensus-based process which aligns with their community-driven approach. For the ODEs studied, this was mainly the role of the platform provider, yet taking the input of the ecosystem, primarily based on their level of importance for the success of the ecosystem and general level of activity. The structure and ownership of the platform provider (considering the cases of organization-centric, consortium-based, and community-based ODEs) further affect how the collective-choice arrangements are made.

\textit{Recommendation \#3 for ODEs:} Rules and overall governance structure should ideally be shaped through dialogue among the ecosystem members, and enable influence, formal or informal, for the members both concerning the governance, as the general collaboration related to the common.

\subsection{Principle 4: Monitoring}

\textit{``Monitors, who actively audit CPR conditions and appropriator behaviour, are accountable to the appropriators or are the appropriators.''}~\cite{Ostrom90}

\subsubsection{Related interpretations on Digital commons:}
Forte et al. interpret the principle much in line with Ostrom as do Vi{\'e}gas et al., stating that \textit{``Individuals who monitor the commons should be accountable to the rest of the community''}. This definition aligns well with observations in the case of Wikipedia where \textit{``[t]he entire community monitors content and if a dispute arises, it is generally resolved through discussion by the people involved in the situation''}~\cite{forte2009decentralization}. In the case of Drupal, Rozas observed the \textit{``[e]mergence of explicit roles and processes related to quality assurance so that certain individuals, accountable to the community, [can] monitor the commons''}~\cite{rozas2017self}.

Ruhaak et al. interpret the principle for data commons and how it \textit{``includes monitoring of data production processes — ongoing validation of data integrity, verification of data quality, — as well as monitoring data access and use''}~\cite{RuhaakMozillaCommons2021}.

\subsubsection{Contextualization on ODEs:}
Monitoring in terms of how rules and governance is followed is not conducted by any external actors but is a responsibility shared between the platform provider and the ecosystem in general. The platform provider may use diagnostics and statistics connected to the platform to monitor how the data is used, as employed by the Swedish Public Employment Service in the case of JobTech. By also mandating use of API keys, they can further monitor e.g., the amount of calls to API:s to a specific user, and require acceptance of any terms and conditions.

From the ecosystem perspective, actors can collectively monitor how the platform provider abides by the defined rules and governance structure, and how they can exercise their influence on e.g., what data is shared.

\subsubsection{Analysis and recommendation:}
In the communities of Wikipedia~\cite{forte2009decentralization} and Drupal~\cite{forte2009decentralization}, the community monitors itself through the transparent and consensus-based collaboration and governance processes. Similar observations regard the ODEs studied. Although, there may not necessarily be equal conditions in terms of monitoring each other, both the platform provider and the members of the ecosystem both have the possibility to monitor and check up on each other, and act accordingly. Traits that are important for a common trust to reside between the members, which is especially central in smaller ecosystems where co-opetition is present, i.e., collaboration between competitors. As Ruhaak et al. point out, monitoring should also include the production and collaboration regarding the common~\cite{RuhaakMozillaCommons2021}.

\textit{Recommendation \#4 for ODEs:} Monitoring concerning the orchestration, as well as the consumption, and general collaboration and production related to the common should be performed equally by both those facilitating an ecosystem (i.e., the platform provider), and the members of the ecosystem.

\subsection{Principle 5: Graduated sanctions}

\textit{``Appropriators who violate operational rules are likely to be assessed graduated sanctions (depending on the seriousness and context of the offence) by other appropriators, by officials accountable to these appropriators, or both.''}~\cite{Ostrom90}

\subsubsection{Related interpretations on Digital commons:}
The interpretation by Forte et al. highlights how \textit{``community members actively monitor and sanction one another when behavior is found to conflict with community rules''.} 
~\cite{forte2009decentralization}

In the Wikipedia community, \textit{``[w]hen behavior-related policy is broken, a series of graduated sanctions can be imposed that begin with the posting of warnings and can lead to being banned from the site''}~\cite{forte2009decentralization}. If the dispute can not be resolved locally in the specific project, the case is transferred and managed by a central entity. In the Drupal community, a code of conduct listing different sanctions has been developed and is enforced collectively by the community. Tweeting derogatory comments about a presenter at a community event may e.g., result in banning from the event~\cite{rozas2017self}.

Coyle et al. emphasize concrete examples of sanctions related to misuse of the data: \textit{``Enforcement of a range of consequences for the misuse of data, ranging from the withdrawal of access permissions to fines and other penalties''}~\cite{CoyleValue2020}. The interpretation by Ruhaak et al. aligns with the original definition~\cite{RuhaakMozillaCommons2021}.

\subsubsection{Contextualization on ODEs:}
As with monitoring, graduated sanctions can be exercised both from the platform provider and ecosystem actors, respectively. The platform provider has the option of disabling or limiting access to the data shared via their platform, especially if they have enforced the use of API keys. In the case of JobTech, API keys have, however, mainly been introduced to communicate any events that might affect the users, e.g., the deprecation or revision of an API that might affect business-critical applications among its users. Another means is to limit the influence that a specific actor may have. 

The ecosystem actors, on the other hand, may enforce informal sanctions by individually or collectively voicing their opinions, or ultimately leaving the ecosystem. This would hurt the platform provider who is dependent on an ecosystem of actors that use and add value to its platform. The actors may, e.g., choose to join an existing ecosystem, or create a new ecosystem, similar to the concept of forking in terms of OSS~\cite{gamalielsson2014sustainability}, a maneuver that is associated with many risks and costs. An alternative sanction is to limit any contribution to the platform and shared resources, e.g., in terms of data -- in our cases alarm data in ESS--CSDL, image data in RoDL or job ads in JobTech.

\subsubsection{Analysis and recommendation:}
As noted in relation to monitoring, definition, and provisioning of sanctions are done collectively by the communities in Wikipedia~\cite{forte2009decentralization} and Drupal~\cite{rozas2017self}. In the ODEs studied, the types of sanctions that can be provided differ between the platform provider and the rest of the ecosystem members. However, relating to Principle \#3, there should be a common discussion and understanding of what sanctions apply and under what conditions. As highlighted by Ruhaak et al.~\cite{RuhaakMozillaCommons2021}, sanctions should consider misconduct both regarding the consumption, and general collaboration and production related to the common.

\textit{Recommendation \#5 for ODEs:} Graduated sanctions should be discussed and decided on collectively by those facilitating an ecosystem (i.e., the platform provider), and the members of the ecosystem. However, different sanctions may still be available to each side even though these should be considered a last resort.

\subsection{Principle 6: Conflict-resolution mechanisms}

\textit{``Appropriators and their officials have rapid access to low cost local arenas to resolve conflicts among appropriators or between appropriators and officials.''}~\cite{Ostrom90}

\subsubsection{Related interpretations on Digital commons:}
The interpretation by Forte et al. aligns with the original definition, as does Vi{\'e}gas et al. who more concisely say that \textit{``community members should have rapid access to low-cost local arenas to resolve conflicts''}. In Wikipedia, most disputes are managed in a decentralized and local setting, a trend that is considered a consequence given the rapid growth of the community~\cite{forte2009decentralization}. In Drupal, there are mechanisms to facilitate conflict resolution which is also managed primarily in the local working groups~\cite{rozas2017self}.

Ruhaak et al. phrase the principle as: \textit{``When conflict arises in a data commons, there needs to be an effective, inexpensive, and otherwise accessible way to handle that conflict. In addition, a data commons needs to decide and make clear which conflicts will be handled internally and which ones should be resolved externally, for instance by going to court''.}~\cite{RuhaakMozillaCommons2021}

\subsubsection{Contextualization on ODEs:}
To enable conflict resolution as well as rulemaking and collaboration in general, there is a need for trust, both between the actors and towards the shared data and its related resources. This can be especially important in ecosystems where there are direct or indirect competitors (as in JobTech and RoDL). Among the cases studied, we have noticed the importance of a neutral platform provider with the possibility to build this trust within the ecosystem. In JobTech, the Swedish Public Employment Service has taken on this role in the context of an organization-centric governance structure. In RoDL and its consortium-based governance structure, an independent body through which multiple actors interact provided a neutral ground for discussions and conflict resolutions.

In ODEs with a community-based governance structure, conflicts are managed through community-defined norms and processes. OpenStreetMap has a consensus-based approach to addressing any disputes, e.g., regarding a certain edit to their map data. \textit{Data trusts} is another solution proposed by Coyle et al.~\cite{CoyleValue2020}, which provides a legal structure that can host any shared data for its users. The OpenStreetMap foundation can be likened to a data trust as it holds the intellectual property rights for the OpenStreetMap community but is not part of the governance structure, as is commonly the case for open source software-focused foundations.

\subsubsection{Analysis and recommendation:}
Both Wikipedia~\cite{forte2009decentralization} and Drupal~\cite{rozas2017self} manage most conflicts in local settings due to their size and decentralized organization, similar to the example of OpenStreetMap. In the ODEs studied, which are considerably smaller in size, conflicts are managed centrally by, or through the platform provider, both in the case of organization-centric and consortium-based ODEs. Ruhaak et al. stress that it should also be defined what conflicts are managed internally and which ones should be managed externally, e.g., in court~\cite{RuhaakMozillaCommons2021} although the latter is somewhat alien to the community context.

\textit{Recommendation \#6 for ODEs:} A neutral actor or common body (potentially the platform provider) can help in establishing a common trust and resolving potential conflicts in the ecosystem. As an ecosystem evolves, a more decentralized process may need to be considered.

\subsection{Principle 7: Minimal Recognition of rights to organize}

\textit{``The rights of appropriators to devise their own institutions are not challenged by external governmental authorities.''}~\cite{Ostrom90}

\subsubsection{Related interpretations on Digital commons:}
Forte et al. reinterpret the principle as: \textit{``Local jurisdiction to create and enforce rules should be recognized by external, central authorities.''}~\cite{forte2009decentralization}
Interestingly, Forte et al. consider this principle in the context of the local projects (or ``jurisdictions'') within the overarching Wikipedia project in contrast to Ostrom et al., who in traditional commons consider the external government and local authorities. In Drupal, there is an \textit{``[e]mergence of local jurisdiction and acknowledgment by the most centralised authorities of creation and enforcement of local rules''}~\cite{rozas2017self}. This development can be observed both in the language-specific communities, as well as among the contributed modules that connect to the Drupal Core OSS project.

Coyle et al. emphasize the need for a \textit{``comprehensive data strategy and institutional/regulatory framework''}~\cite{CoyleValue2020}. In the context of data commons, as per Ruhaak et al., \textit{``this principle encourages us to understand how far the decisions we make about the collection and use of data are in line with, for instance, data protection regulations''}~\cite{RuhaakMozillaCommons2021}.

\subsubsection{Contextualization on ODEs:}
An ODE needs legitimacy to attract actors and stay competitive with other ecosystems. The legitimacy, as well as attractiveness, relates to the value an ODE can offer its actors, e.g., through the data being shared, and the related collaboration. For ODEs focused on open government data, public organizations (as with JobTech) can provide the legitimacy and foundation for the trust needed to establish a sustainable and long-term ecosystem. An ODE must also respect regulations and aspects that might limit the kind of data that may be shared, or how.

\subsubsection{Analysis and recommendation:}
This principle is contextualized both in the case of Wikipedia~\cite{forte2009decentralization} and Drupal~\cite{rozas2017self} inside each community. This comes naturally given the size and decentralized nature of the communities, creating the dynamics and need for respect between central and local governance. The contexts of the studied ODEs align more with the original definition by Ostrom in terms of a need to be recognized by the external environment (not necessarily local authorities as stated by Ostrom). Such recognition may be built by including necessary parties as members of the ecosystem, e.g., public sector organizations or actors who could be considered potential keystones. As highlighted by Ruhaak et al., one also needs to consider the relevant regulations, e.g., in terms of data protection regulations~\cite{RuhaakMozillaCommons2021}. Further, as highlighted by Principle \#2, an ecosystem also needs to consider risks perceived by the members and how these can be addressed.

\textit{Recommendation \#7 for ODEs:} External recognition is connected to the attractiveness of the ecosystem and its underpinning common, and the value they can provide to new and existing members. The ecosystem must also ensure that regulations are met and perceived risks from its members are addressed as highlighted in Principle \#2.

\subsection{Principle 8: Nested enterprises (for CPRs that are part of larger systems)}

\textit{``Appropriation, provision, monitoring, enforcement, conflict resolution, and governance activities are organized in multiple layers of nested enterprises.''}~\cite{Ostrom90}

\subsubsection{Related interpretations on Digital commons:}
This principle only relates to CPRs that are part of the larger ecosystem and overarching governance. Forte et al. define it as: \textit{``By forming multiple nested layers of organization, communities can address issues that affect resource management differently at broader and very local levels.''}~\cite{forte2009decentralization}

Again, this principle can be contextualized in the context of the overarching Wikipedia community, and how its governance structure considers the local projects within~\cite{forte2009decentralization}. In Drupal, Rozas noted an \textit{``[o]verall emergence of socio-technical systems of contribution to address issues that affect the common resources differently at wider and local levels''}, e.g., through the emergence of international, regional, and local conferences and meetups where members of the community and concerned sub-communities can meet and discuss matters of importance~\cite{rozas2017self}.

Concerning data commons, \textit{``this principle can refer to a possible need for one data commons to interoperate with another, or to break up one large commons into smaller, nested commons that interoperate with one another. Doing so would allow each smaller commons to make decisions that better reflect their circumstance and match a narrowly defined purpose''}~\cite{RuhaakMozillaCommons2021}.

\subsubsection{Contextualization on ODEs:}
It is not uncommon that an ODE is part of a larger and more nested ecosystem. In the case of JobTech, the taxonomy of skill and work titles on one hand relates to other data sets such as those defining types of education maintained by the National Agency for Education. On another hand, the taxonomy data set makes up a Swedish translation and contextualization of a more abstract taxonomy data set for the European Union, which in turn relates to an even more abstract international taxonomy.



\subsubsection{Analysis and recommendation:}
In both the Wikipedia~\cite{forte2009decentralization} and the Drupal communities~\cite{rozas2017self}, this principle is contextualized as the different layers that emerge due to their large size and decentralized nature. Again, for the ODEs studied, which are all considerably smaller in size, this principle rather refers to how an ODE relates to connecting ODEs, on a similar or differing level of abstraction as with the example of public transport data. This aspect is also highlighted by Ruhaak et al.~\cite{RuhaakMozillaCommons2021} as the potential need to interoperate with other ODEs, or even break one up into smaller ones. In some sense, similar to what has happened in Wikipedia and Drupal, where smaller sub-communities have emerged forming new and lower levels within the overarching community. An ODE may, hence, as it matures turn into a nested enterprise itself with layers, and local jurisdictions emerging underneath. 

\textit{Recommendation \#8 for ODEs:} Interoperability and collaboration should be sought out by connecting ecosystems with aligning commons. Ecosystems should further be open to internal sub-groupings emerging as the ecosystem evolves, or even breaking up an ecosystem if appropriate, e.g., enabling smaller commons with a narrower focus.

\section{Discussion and Conclusions}



\subsection{Natural and digital commons}
When reanalyzing the ODE cases as digital commons~\cite{morell2010governance}, in relation to natural commons, one difference sticks out, namely whether the resource is subtractable as hinted by Ostrom's definition of a CPR~\cite{ostrom2000collective}. A subtractable resource is reduced through the consumption by one on the cost of others. If non-renewable, the resource may risk depletion. A non-subtractable resource, on the other hand, can be consumed by multiple stakeholders without being diminished.

For natural commons, e.g water supply in a well or creek, the dilemma of the commons is clearly prevalent. If one landowner along the creek taps too much water for irrigation, the creek will dry out for the downstream land owners. The same phenomenon is observable at multiple levels, from the small village well, to huge-scale power dam projects, e.g. in Ethiopia~\cite{GrandDam}. Consequently, commons must focus on regulating consumption, and sanctions taken against those who break the regulations. Natural commons resources may also suffer from under-consumption, e.g. if too few creatures graze a common pasture, bushes may take over the land or the ground may be malnourished due to lack of fertilization from the creep if too small herds use the pastures. However, the primary concern is over-consumption.

For digital commons, e.g. OSS~\cite{schweik2012internet, jullien2020digital} and open data~\cite{CoyleValue2020, RuhaakMozillaCommons2021}, the dilemma of the commons is different. Digital commons are primarily non-subtractable. Software is not worn out by being run, and data is not consumed by being copied to new databases. On the contrary, OSS and open data are improved by utilization. More users of an OSS means that more eyes spot potential faults, and more stakeholders have an interest in having these faults fixed. Thus the probability that some stakeholders actually will provide labor to fix the fault is higher. For open data, similar patterns can be seen, e.g. when users of OpenStreetMap spot faults in the map data and propose corrections. Further, they map the unchartered territory and make this data available to other consumers.

However, this highlights another aspect of digital commons that is very much subtractable. That is the labor needed for the collective action to properly maintain the digital common. For example, if digital preservation efforts such as the Software Heritage~\cite{di2017software} would fail if a lack in preservation efforts arises. Equally, if an OSS is not maintained properly by the community, vulnerabilities may risk being introduced and dependencies break. In the case of open data, ML training sets may risk becoming invalid, or map data incorrect as its physical counterpart is altered. 

As a community grows, and so the appropriation of a common, the demand grows for maintenance labor that can provide support for the community, e.g., by answering questions~\cite{eghbal2016roads}. Effort required to manage and coordinate collective action for maintaining the common may also risk to scale out of proportion. Or as put by Jullien and Roudaut, \textit{``the available brain time of peers is not extensible, therefore rival, even if it is a renewable resource''}~\cite{jullien2020digital}. The collective action producing and maintaining the common, may in this sense itself be considered as a CPR where the ``brain time'' makes up the resource system that needs governing and oversight. 

Communities, hence, need to design their governance in a way that manages a potential increase in effort as appropriation and provisioning grows (as in the examples of Wikipedia~\cite{forte2009decentralization, viegas2007hidden} and Drupal~\cite{rozas2017self}), while also reducing barriers to entry and enabling new members of a community to engage and start contributing to the collective action as frictionless as possible~\cite{jullien2020digital}. We synthesize our findings into two propositions:

\textbf{P1}) As appropriation of a digital common grows, so does the demand for labor required to contribute to and manage the contributions to the collective action, underpinning the common.

\textbf{P2}) To enable sustainability of a digital common, governance must consider both how to promote and manage contributions, within the available frame of maintenance labor of a community.

\subsection{Design principles for digital commons}

As a consequence of the difference between natural and digital resources, we have tailored some of Ostrom's design principles conceptually, while others only have to be contextualized into the digital domain.

Two principles remain principally the same for digital commons, \#1 defined boundaries, and \#3 collective-choice arrangements. In both natural and digital commons, there is a need for clear scope definitions and the democratic aspects of influencing the common. The principles may be implemented differently: the boundaries of the artificially created artifacts of digital commons are scoped by their vision, while boundaries of the natural commons are related to physical entities and stakeholders~\cite{benkler2014between}. Similarly, the mechanisms for enabling influence from stakeholders are different in the digital and the physical world. Still, in both cases, the dialogue, not its medium is the key.

Another three of the principles are also principally the same, but since the digital media are so different, the principles are conceptually modified with respect to what is local and what is global. For principles \#6, conflict resolution, \#7, rights to organize, and \#8, nested enterprises, the commons are technically global (via the internet) but there is little global-wide jurisdiction to align to. Thus, local working groups are created e.g. within the global Wikipedia community, to decentralize the governance~\cite{forte2009decentralization, viegas2007hidden}. This is in contrast to natural commons, where the scope of the common is physically bound to a country or a region~\cite{hess2008mapping}. Principally, also natural commons are global, e.g. $CO_2$ and water resources, but we have not seen commons principles applied at this scale.

The remaining three principles are conceptually different. Principle \#2 on appropriation and provision is for natural commons focused on over-consumption, while governance of digital commons must monitor under-consumption and ensure the incentives for new contributions to the commons, whether it is software or data. 
For principles \#4 on monitoring and \#5 on sanctions, the platform provider in the digital common has a central role, since monitoring and sanctions are technically connected to features of the platform. The APIs of the platform may handle access rights, and sanctions can be implemented as withdrawn API keys, giving the platform provider's powerful tools to manage the common~\cite{LinakerJeDEM21}. However, since the stakeholders are needed for new contributions to the common, this helps balance the power between the facilitators (platform providers) and members (users) of the common.

\subsection{Limitations and Future work}


The applicability of Ostrom's design principles to digital commons may be questioned as they were originally elicited for small and locally based natural commons. However, as proposed by Benkler~\cite{benkler2014between}, and explored by others~\cite{viegas2007hidden, forte2009decentralization, rozas2017self, CoyleValue2020, RuhaakMozillaCommons2021}, they may very well prove applicable to digital commons, although with some modifications as hinted by this study.


A further aspect concerns when open data can be considered a digital common, and by extension, to which ODEs that our recommendations are applicable. Many authors (e.g.,~\cite{morell2010governance, benkler2014between, jullien2020digital}) highlight the characteristics of co-creation and co-ownership. A (hypothetical) case of an organization-centric business-driven ODE where a company release open data as part of their business model for public consumption may hence be questioned. However, per our definition of an ODE, if there is no collaborative component, or networked group of actors present with a common vision (see Section~\ref{subsect:rw:OGD}), it is questionable if this hypothetical case can be considered as an ODE. Our recommendations may instead be considered as guidance for how such a company can organize an ODE and tailor its governance structure to transform the open data into (or resemble aspects of) a ``digital common'' and thereby extract the potential benefits that (may) follow. Research into the gray area of what constitutes a digital common is a topic warranting further attention.

It should further be noted that our review of related interpretations of Ostrom design principles is limited to the referred works~\cite{forte2009decentralization, viegas2007hidden, rozas2017self}, and their application of the principles when describing the governance of their respective cases. We hence acknowledge that our coverage of, e.g., social and cultural aspects that also affect the governance, i.e., the social norms practiced, are not extensively covered, and is considered as a topic for future research.


\section*{Acknowledgement}
Thanks to professor Krister Andersson, University of Colorado at Boulder, for advice and encouragement in applying the commons concept to the digital domain.  

\bibliographystyle{ACM-Reference-Format}
\bibliography{references}
\end{document}